\documentclass[runningheads]{assets/llncs}
\usepackage{graphicx}
\usepackage{booktabs}
\usepackage{graphicx}
\usepackage{wrapfig}
\usepackage{amsmath}
\usepackage{amsfonts}
\usepackage{amssymb}
\usepackage{array}
\usepackage{adjustbox}
\usepackage{multirow}
\usepackage{colortbl}
\usepackage{caption}
\usepackage{amsmath}
\usepackage{multicol}
\usepackage{mathtools}
\usepackage{array,multirow}
\usepackage{makecell}
\usepackage{tikz}
\usepackage{xparse}

\usepackage{soul}

\usepackage{color,xcolor}

\usepackage{url}
\usepackage{algorithm, algorithmic}
\usepackage[super]{nth}

\makeatletter
\DeclareRobustCommand\onedot{\futurelet\@let@token\@onedot}
\def\@onedot{\ifx\@let@token.\else.\null\fi\xspace}

\makeatother

\newcommand{\R}{\mathbb{R}}

\newcommand{\D}[0]{\mathcal D}

\NewDocumentCommand\DownArrow{O{2.0ex} O{black}}{%
   \mathrel{\tikz[baseline] \draw [<-, line width=0.5pt, #2] (0,0) -- ++(0,#1);}
}
\NewDocumentCommand\UpArrow{O{2.0ex} O{black}}{%
   \mathrel{\tikz[baseline] \draw [<-, line width=0.5pt, #2] (0,0) -- ++(0,#1);}
}

\definecolor{rowblue}{RGB}{220,230,240}
\definecolor{myorchid}{RGB}{150,10,30}
\definecolor{myblue}{RGB}{10,30,250}
\definecolor{mygreen}{RGB}{10,120,10}

\begin{document}
\title{Automatic Segmentation of the Placenta in BOLD MRI Time Series}
\titlerunning{Automatic Segmentation of the Placenta in BOLD MRI}
%
 \author{Anonymous}
\author{S. Mazdak Abulnaga\inst{1} \and
Sean I. Young\inst{1,2} \and
Katherine Hobgood\inst{1} \and
Eileen Pan\inst{1} \and
Clinton J. Wang\inst{1} \and
P. Ellen Grant\inst{3} \and
Esra Abaci Turk\inst{3} \and
Polina Golland\inst{1}
}
%
\authorrunning{Abulnaga et al.}
%

\institute{Computer Science and Artificial Intelligence Lab, Massachusetts Institute of Technology, Cambridge, 02139, USA  \and
MGH/HST Martinos Center for Biomedical Imaging, Harvard Medical School, Boston, MA, 02129, USA \and
Fetal-Neonatal Neuroimaging and Developmental Science  Center, Boston Children's Hospital, Harvard Medical School, Boston, MA, 02115, USA \\
\email{abulnaga@mit.edu, siyoung@mit.edu, khobgood@mit.edu, eileenp@mit.edu,}\\
\email{clintonw@csail.mit.edu, ellen.grant@childrens.harvard.edu,} \\
\email{esra.abaciturk@childrens.harvard.edu, polina@csail.mit.edu}
}
\maketitle
\begin{abstract}
Blood oxygen level dependent (BOLD) MRI with maternal hyperoxia can assess oxygen transport within the placenta and has emerged as a promising tool to study placental function. Measuring signal changes over time requires segmenting the placenta in each volume of the time series. Due to the large number of volumes in the BOLD time series, existing studies rely on registration to map all volumes to a manually segmented template. As the placenta can undergo large deformation due to fetal motion, 
maternal motion, and contractions, this approach often results in a large number of discarded volumes, where the registration approach fails. In this work, we propose a machine learning model based on a U-Net neural network architecture to automatically segment the placenta in BOLD MRI and apply it to segmenting each volume in a time series. We use a boundary-weighted loss function to accurately capture the placental shape. Our model is trained and tested on a cohort of $91$ subjects containing healthy fetuses, fetuses with fetal growth restriction, and mothers with high BMI. We achieve a Dice score of $0.83\pm0.04$ when matching with ground truth labels and our model performs reliably in segmenting volumes in both normoxic and hyperoxic points in the BOLD time series. Our code and trained model are available at \url{https://github.com/mabulnaga/automatic-placenta-segmentation}.


\end{abstract}

\section{Introduction}
The placenta is an organ that provides oxygen and nutrients to support fetal growth. Placental dysfunction can cause pregnancy complications and can affect fetal development, so there is a critical need to assess placental function \textit{in vivo}.
Blood oxygen level dependent (BOLD) MRI can directly quantify oxygen transport within the placenta~\cite{sorensen2013changes,turk2019placental} and has emerged as a promising tool to study placental function. Temporal analysis of BOLD MRI with maternal oxygenation has been used to identify contractions~\cite{abaciturk2020placenta,sinding2016reduced}, biomarkers of fetal growth restriction~\cite{luo2017vivo,sorensen2015placental}, predict placental age~\cite{pietsch2021applause} and to study congenital heart disease~\cite{you2020hemodynamic,steinweg2021t2} among many uses.

\begin{figure}[t]
    \includegraphics[width=\linewidth]{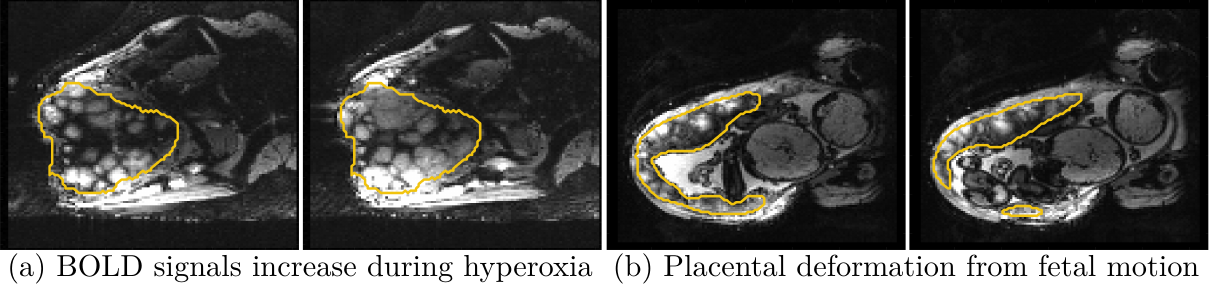}
    \caption{Example images and placental segmentations: (a) signal brightening during hyperoxia, and (b) shape deformation caused by fetal motion. Placental boundaries are marked in yellow. Areas outside of the placenta are darkened for illustration. Intensity scale is based on the first MRI volume in the time series.}
    \label{fig:BOLD-example}
\end{figure}

Despite its importance for many downstream clinical research tasks, placental segmentation is often performed manually and can take a significant amount of time, even for a trained expert. For BOLD MRI studies, manual segmentation is rendered more challenging due to the sheer number of MRI scans acquired and rapid signal changes due to the experimental design. Experiments acquire several hundred whole-uterus MRI scans to observe signal changes in three stages: i) normoxic (baseline), ii) hyperoxic, and iii) return to normoxic. During the hyperoxic stage, the BOLD signals increase rapidly, leading to hyperintensity throughout the placenta. Furthermore, the placental shape can undergo large deformation caused by maternal breathing, contractions, and fetal motion which can be particularly increased during hyperoxia~\cite{you2015robust}. See Fig.~\ref{fig:BOLD-example} for two examples.

The current practice is to analyze BOLD signals with respect to one template volume. Deformable registration of all volumes in the time series to the template is performed to enable spatiotemporal analysis~\cite{turk2017spatiotemporal,you2015robust}. However, due to significant motion, registration can lead to large errors, requiring outlier detection and possibly rejecting a significant number of volumes~\cite{turk2017spatiotemporal,you2015robust}. 

To address these challenges, we propose a model to automatically segment the placenta in BOLD MRI time series. Our model is trained on several volumes from each patient during the normoxic and hyperoxic phases, to capture the nuanced placental changes. We apply our model on unseen BOLD MRI volumes to demonstrate consistency in the predicted segmentation label maps. Our method performs favorably against the state-of-the-art on a large dataset with a broad range of gestational ages and pregnancy conditions. Automatic segmentation is necessary for whole-organ signal analysis, and can be used to improve time-series registration to enable localized analysis. Furthermore, it is an essential step in several post-processing tasks, including motion correction~\cite{turk2017spatiotemporal}, reconstruction~\cite{uus2020deformable}, and mapping to a standardized representation~\cite{miao2017placenta,abulnaga2021volumetric}.

Machine learning segmentation models for the placenta have been previously proposed and include both semi-automatic~\cite{wang2015slic} and automatic~\cite{alansary2016fast,torrents2019fully,pietsch2021applause,specktor2021bootstrap} approaches. While semi-automatic methods have achieved success in predicting segmentation label maps with high accuracy, these approaches are infeasible for segmenting BOLD MRI time series due to the large number of volumes. The majority of automatic methods focus on segmentation in anatomical images. Alansary et al.~\cite{alansary2016fast} proposed a model for segmenting T2-weighted (T2w) images based on a 3D CNN followed by a dense CRF for segmentation refinement and validated on a singleton cohort that included patients with fetal growth restriction (FGR). Torrents-Barrena et al.~\cite{torrents2019fully} proposed a model based on super-resolution and an SVM and validated on a singleton and twin cohort of T2w MRI. Spektor-Fadida et al.~\cite{specktor2021bootstrap} tackled the problem of domain transfer by a self-training model and demonstrated successful segmentation of FIESTA and TRUFI sequences. For a more detailed treatment of segmentation methods in fetal MRI, we refer the reader to the survey by Torrents-Barrena et al.~\cite{torrents2019segmentationsurvey}.

Functional images of the placenta differ greatly from anatomical images, as they have lower in-plane resolution and the contrast between the placental boundary and surrounding anatomy is less pronounced. Anatomical images may also benefit from super-resolution approaches to increase SNR in the acquired image~\cite{uus2020deformable}. Pietsch et al.~\cite{pietsch2021applause} are the first to consider placental segmentation in functional MRI. They proposed a 2D patch-based U-Net model for functional image segmentation and demonstrated a successful application of age prediction using the estimated T2* values. They focused on a cohort of singleton subjects, and demonstrated success on abnormal pregnancy conditions including preeclampsia. In contrast to their approach that segments derived T2* maps, we evaluate our segmentation model on BOLD MRI time series. Furthermore, our 3D model operates on the entire volume rather than patches, thereby helping to better resolve the boundaries of the placenta.

To capture the large signal changes and placental shape variation in the time series, we train with a random sampling of manual segmentations of several volumes in the BOLD MRI series. We propose a boundary weighted loss function to more easily identify the placental boundary and improve segmentation accuracy. Finally, to evaluate the feasibility of our method for clinical research, we propose additional metrics to evaluate performance on the whole MRI time series, and illustrate a possible clinical research application.

\section{Methods}

We aim to find a model $F_{\theta}:X\rightarrow Y$ that takes a BOLD MRI time series $X \in \R^{T\times H \times W \times D}$ and predicts a set of placenta segmentation label maps for each time point $t \in \{1,\dots T\}$, $Y\in \{0,1\}^{T \times H \times W \times D}$, where $T$ is the total number of time points at which MRI scans were acquired. For a given BOLD time series, we have a small number $N_l$ of frames with ground truth labels $(\mathbf x, \mathbf y)$, where $\mathbf x \in \R^{H \times W \times D}$ is an MRI scan and $\mathbf y \in \{0,1\}^{H \times W \times D}$ is the ground truth placenta label map.  
\subsection{Model}
\begin{figure}[t]
\centering
    \includegraphics[width=0.85\linewidth]{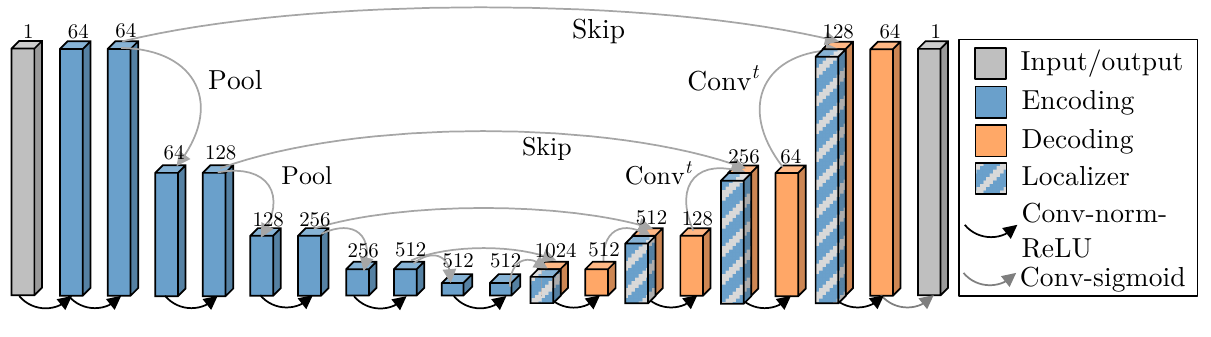} 
    \caption{3D Placenta Segmentation U-Net. We use a five-level 3D U-Net with max-pooling, skip connections, and convolution-transpose layers. Numbers above vertical bars denote the number of features at various stages of the processing pipeline. Batch norm is employed for normalization ($\textrm{batch size} = 8$).}
    \label{fig:U-Net}
\end{figure}

We use a 3D U-Net~\cite{ronneberger2015unet} with $4$ blocks in the contracting and expanding paths. Each block consists of two sets of $3\times 3 \times 3$ convolution with ReLU activations, followed by max pooling (contraction path) or transpose convolution (expansion path), as illustrated in Fig.~\ref{fig:U-Net}. We augment the images using random affine transforms, flips, whole-image brightness shifts, contrast changes, random noise, and elastic deformations, using TorchIO~\cite{garcia2021torchio}. We simulate the effects of maternal normoxia and hyperoxia with a constant intensity shift in the placenta.

To capture the MRI signal and placental shape changes resulting from maternal hyperoxia and fetal motion, we enhance our training with several manually segmented volumes in the normoxic or hyperoxic phase. This allows the model to learn from the realistic variations that arise during maternal oxygenation.

\subsection{Additive Boundary Loss}
The placental boundary can be difficult to distinguish in BOLD MRI scans due to similar appearance with surrounding anatomy. To emphasize the boundary details, we construct an additive boundary-weighting $W$ to the segmentation loss function $L$. Given a ground truth placental label map $\mathbf y$, we denote its boundary as  $\partial \mathbf y$. We use a signed distance function $f(x)$ that measures the signed distance, $d(x, \partial \mathbf y)$, of voxel $x \in \R^3$ to the boundary, where $f(x)<0$ when outside of the placenta and $f(x)>0$ when inside. The boundary weighting is  additive for voxels within $\delta$-distance of $\partial \mathbf y$,

\begin{equation}
\label{eqn-boundary-weighting}
W_{\delta }(x) =
\begin{cases}
w_1 & \text{if } -\delta<f(x)<0, \\
w_2 & \text{if } \quad 0 \leq f(x)<\delta, \\
0 & \text{otherwise}.
\end{cases}
\end{equation}
The weighted-loss is then
\begin{equation}
\label{eqn-boundary-loss}
L_w\left(x \right) = L\left(x \right)\left[1+W_{\delta}\left(x\right)\right].
\end{equation}
 In practice, we set $w_1>w_2$, to penalize outside voxels more heavily and learn to distinguish the placenta from its surrounding anatomy. To find voxels with $|f(x)|<\delta$, we estimate a $2\delta$-wide boundary by an average pooling filter on $\mathbf y$ with kernel size $K$ and take the smoothed outputs to lie in the boundary. A larger $K$ produces a wider boundary, penalizing more misclassified voxels. 

\subsection{Implementation Details}

We train using a learning rate $\eta = 10^{-4}$ for $3000$ epochs and select the model with the best Dice score on the validation set. For the additive boundary loss, we set $w_1=40$, $w_2=1$, and $K=11$. All volumes are normalized by mapping the \nth{90} percentile intensity value to $1$. We use a batch size of $8$ MRI volumes. We crop or pad all volumes in the dataset to have dimension $112\times112\times80$, and train on the entire 3D volume. We augment our data with random translations of up to 10 voxels, rotations up to 22$^\circ$, Gaussian noise sampled with $\mu=0,\sigma=0.25$, elastic deformations with $5$ control points and a maximum displacement of $10$ voxels, whole volume intensity shifts up to $\pm 25\%$, and whole-placenta intensity shifts of $\pm0.15$ normalized intensity values. These values were determined by cross-validation on the training set. When evaluating the model on our test set, we post-processed produced label maps by taking the largest connected component to eliminate islands. Our code and trained model are available at \url{https://github.com/mabulnaga/automatic-placenta-segmentation}.  

\section{Model Evaluation}

\subsection{Data}
Our dataset consists of BOLD MRI scans taken from two clinical research studies. Data was collected from $91$ subjects of which $78$ were singleton pregnancies (gestational age (GA) at MRI scan of $23$wk$5$d -- $37$wk$6$d), 
and $13$ were monochorionic-diamniotic (Mo-Di) twins (GA at MRI scan of $27$wk$5$d -- $34$wk$5$d). Of these, $63$ were controls, $16$ had fetal growth restriction (FGR), 
and $12$ had high BMI (BMI $>30$). Obstetrical ultrasound was used to classify subjects with FGR. For singleton subjects, classification was done based on having fetuses with estimated weight less than the \nth{10} percentile. For twin subjects, FGR classification was determined by provene monoochorionicity and discordance in the estimated fetal weight by i) growth restrction ($<$\nth{10} percentile) in one or both fetuses; and/or ii) growth discordance ($\geq 20\%$) between fetuses. Table~\ref{tab:subject-demo} shows patient demographics and GA ranges per group.

\begin{table}[t]
\caption{Subject demographic information.}
\label{tab:subject-demo}
\centering
    \includegraphics[width=0.75\linewidth]{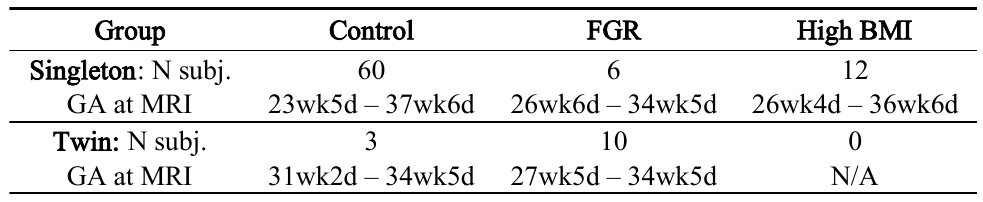}
\end{table}

MRI BOLD scans were acquired on a $3$T Siemens Skyra scanner (GRE-EPI, interleaved with $3$mm isotropic voxels, TR $=5.8$--$8$s, TE $=32-47$ ms, FA = $90^{\circ})$. To eliminate intra-volume motion artifacts, we split the acquired interleaved volumes into two separate volumes with spacing $3\times 3 \times 6$mm, then linearly interpolate to  $3 \times 3 \times 3$mm. In our analysis, we only consider one of two split volumes.
Maternal oxygen supply was alternated during the BOLD acquisition via a nonrebreathing facial mask to have three 10-minute or 5-minute consecutive episodes: 1. Normoxic (21\% $O_2$), 2. Hyperoxic (100\% $O_2$, 15L/min), 3. Normoxic (21\% $O_2$). The placenta was manually segmented by a trained observer. Each BOLD MRI time series had $1$ to $6$ manual segmentations, yielding a total of $176$ ground truth labels. The data was split into a training, validation, and test sets: ($65\%/15\%/20\%$: $63/11/17$ subjects) and stratified on pregnancy condition.

 Each subject in the training set had up to $N_l=6$ ground truth segmentations in the BOLD time series. To prevent the model from being biased by subjects with more ground truth labels,  we train by randomly sampling one of $N_l$ ground truth segmentations in each epoch. 





\subsection{Evaluation}We first compare the predicted segmentation label maps to ground truth segmentations. We measure similarity using the Dice score (Dice), the \nth{95}-percentile Hausdorff distance (HD95), and the Average Symmetric Surface Distance (ASSD). To evaluate the feasibility of the produced segmentations for clinical research studying whole-organ signal changes, we evaluate the relative error in the mean BOLD values, defined as $\vert\hat{b}-b\vert/{b}$, where $b$ and $\hat{b}$ denote the mean BOLD signal in the ground truth and in the predicted segmentation, respectively. 

We evaluate several variants of our model using these metrics. We assess the effect of the boundary-weighting (BW) loss term and compare performance using the Cross-entropy (CE), Dice~\cite{Milletari2016VNetFC}, and Focal~\cite{lin2017focal} loss functions. We evaluate the generalization ability by comparing with the model trained on only the first of $N_l$ BOLD frames and without random sampling of labeled segmentations.

We evaluate our model's sensitivity to oxygenation by comparing the accuracy of predictions in the normoxic and hyperoxic phases for a given subject. We compute the absolute difference of the similarity metric $m$ between an image in normoxia and in hyperoxia, $\vert m_{normoxic}(\mathbf y, \hat{\mathbf y} )-m_{hyperoxic}(\mathbf y, \hat{\mathbf y}) \vert$, where $m_{normoxic}(\mathbf y, \hat{\mathbf y})$ denotes the similarity between our predicted segmentation $\hat{\mathbf y}$ and the ground truth $\mathbf y$ using the metric $m$ for an image in the normoxic phase. We use the Dice score, HD95, ASSD, and relative BOLD error for $m$.

We assess the consistency of our predictions by applying our model to all volumes in the BOLD time series of the test set. Since our volumes are acquired interleaved and split into two separate volumes, we apply our model to every second volume in the time series, yielding a mean of $111.7\pm45.3$  volumes per subject. We measure consistency by comparing the Dice score, HD95, ASSD, and normalized BOLD difference between consecutive volumes. 

Finally, we demonstrate a possible application of temporal analysis by measuring increases in mean BOLD signal during hyperoxia.

\subsection{Results}

\begin{table}[t]
\caption{Test results produced by our 3D U-Net model trained using different loss functions. Numbers in bold indicate the best result in each column.}
\label{tab:test-results}
   \includegraphics[width=\linewidth]{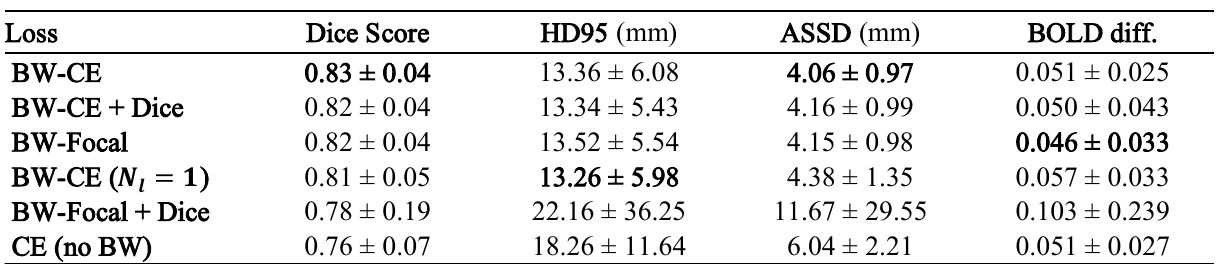}
\end{table}
Table~\ref{tab:test-results} reports the performance of several variants of our model on the test set. Our best model achieves a Dice score of $0.83\pm 0.04$ with a HD$95=13.36\pm6.08$mm using the BW-CE loss. Further, we achieve low relative BOLD error ($0.051\pm 0.025$), indicating that our model's segmentations are suitable for clinical research studies assessing whole-organ signal changes. Similar performance is achieved for the other loss functions. Training the model without the boundary weighting (Eq.~\eqref{eqn-boundary-loss}) results in a statistically significant drop in performance, achieving a Dice of $0.76$ ($p<10^{-4}$ using a paired t-test). Using only the first segmented volume of the BOLD MRI series ($N_l=1$) in the normoxic phase also results in a significant drop in performance, achieving a Dice of $0.81$ ($p<0.05$). Adding labeled examples in the hyperoxic phase helps generalization, as the placental shape and intensity patterns can change greatly. 


Our performance is consistent across pregnancy conditions, as we achieve Dice scores of ($0.76$, $0.89$) on the two subjects with twin pregnancies, $0.83\pm0.04$ on the singletons (N$=15$), $0.83\pm0.07$ on the FGR cohort (N=$3$), $0.82\pm0.04$ on the controls (N=$12$) and $(0.84,0.88)$ on the two BMI cases.  

Direct comparison of this work to previous studies is not feasible due to differences in data set size and patient demographics, imaging protocols, and MRI study design. The current state-of-the-art automatic segmentation method for functional MRI (T2*) achieves a Dice score of $0.58$ on a cohort of low- and high-risk singleton subjects of a wide GA range~\cite{pietsch2021applause}. Their performance was comparable to the inter-rater variability of two radiologists (Dice=$0.68$), which represents an upper limit. In their work, they trained on a combination of T2* weighting and BOLD sequences, while we focus only on BOLD.

\begin{figure}[t]
\scriptsize
    \centering
    \begin{tabular}{ccccc}
    \includegraphics[width=.19\linewidth]{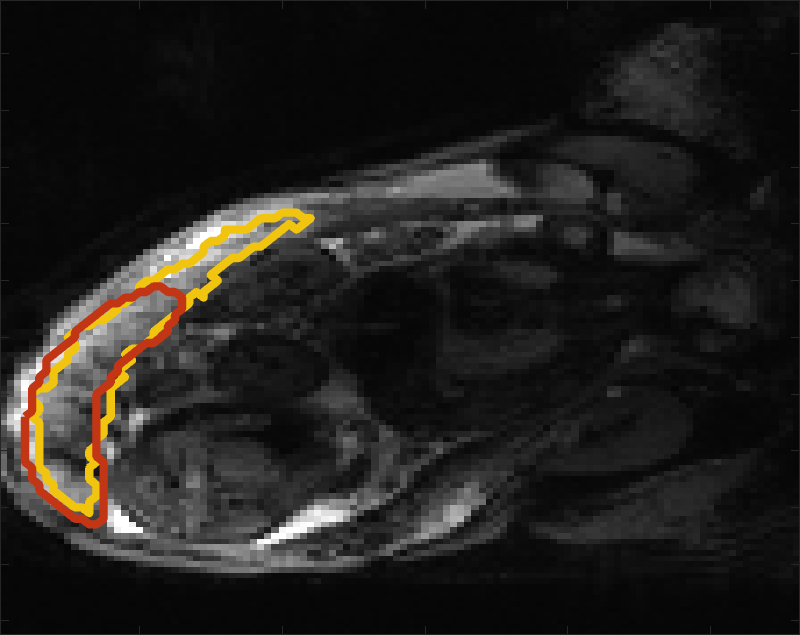} &
    \includegraphics[width=.19\linewidth]{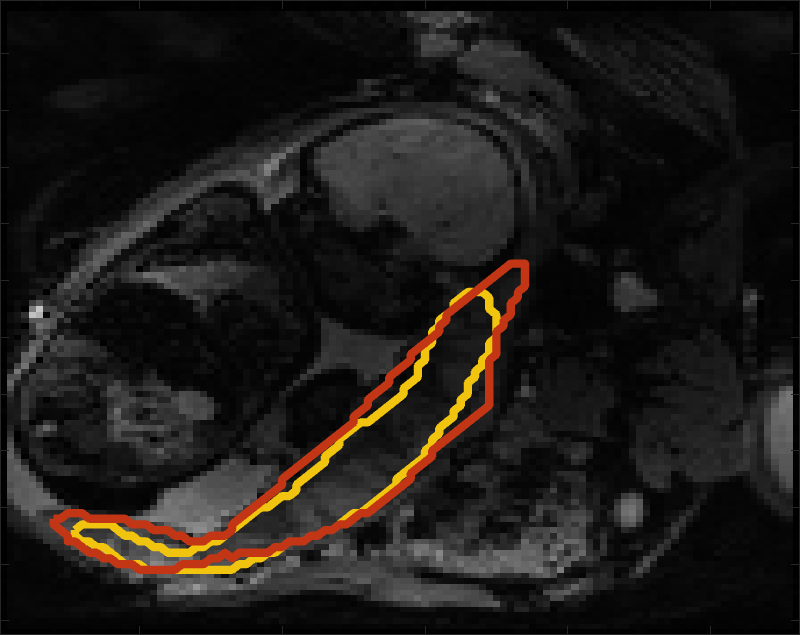} &
    \includegraphics[width=.19\linewidth]{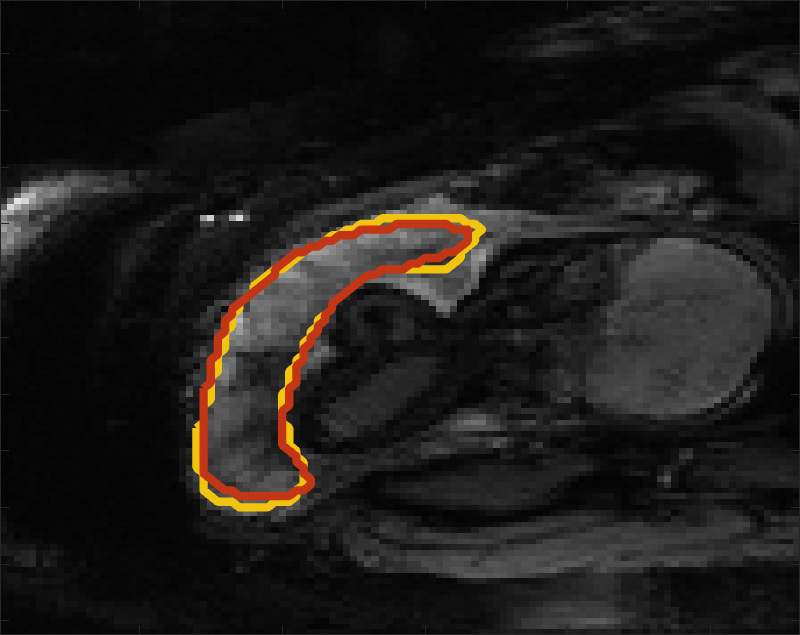} &
    \includegraphics[width=.19\linewidth]{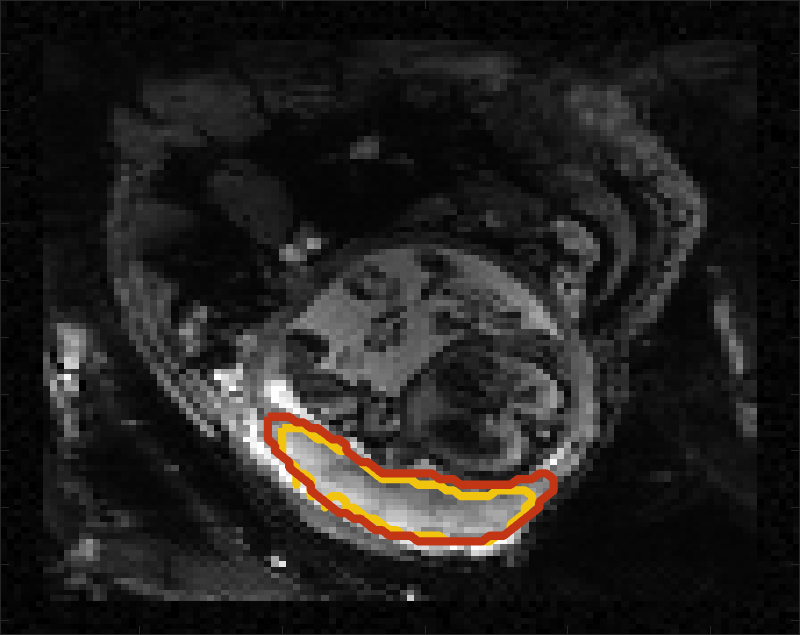} & \includegraphics[width=.19\linewidth]{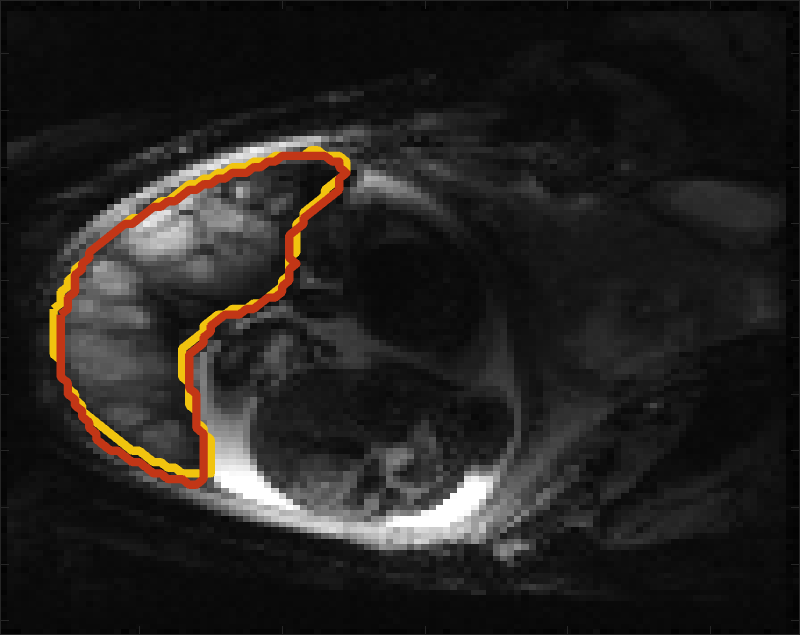} \\
    
    \includegraphics[width=.19\linewidth]{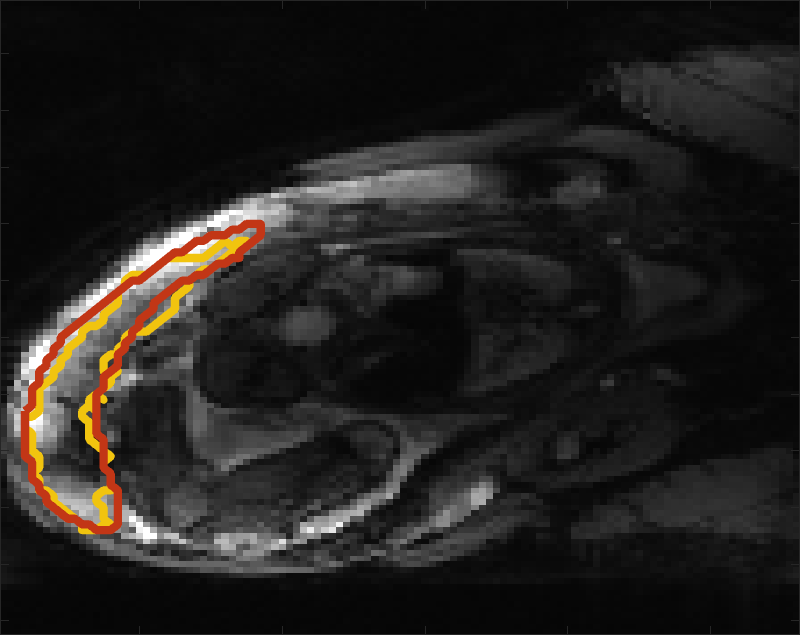} &
    \includegraphics[width=.19\linewidth]{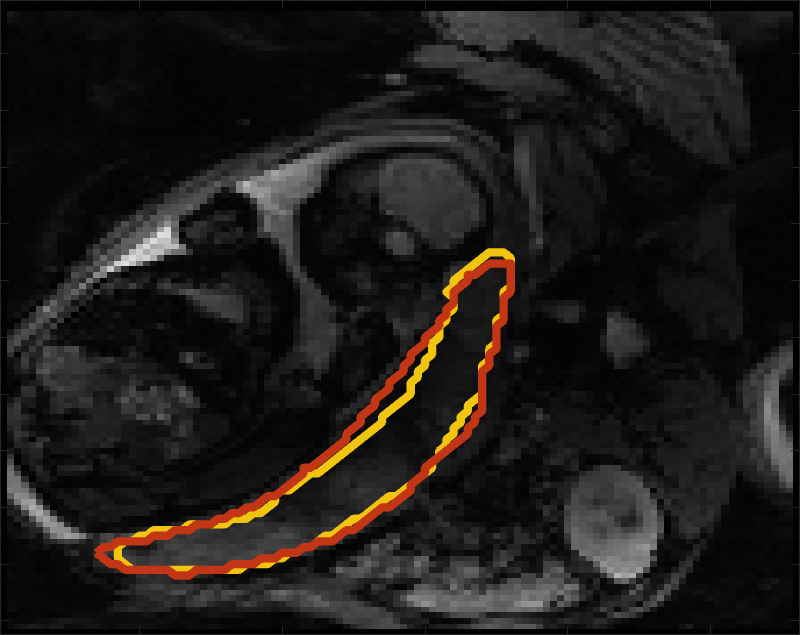} &
    \includegraphics[width=.19\linewidth]{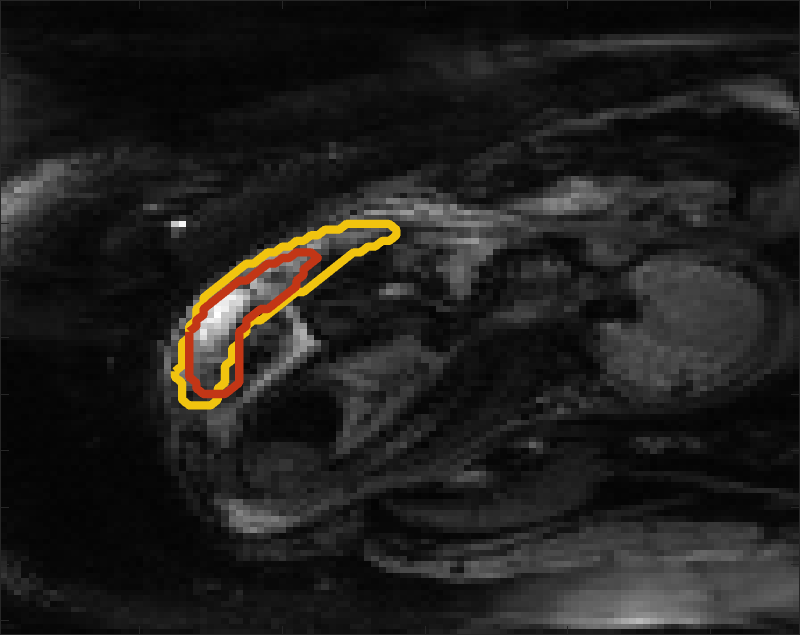} &
    \includegraphics[width=.19\linewidth]{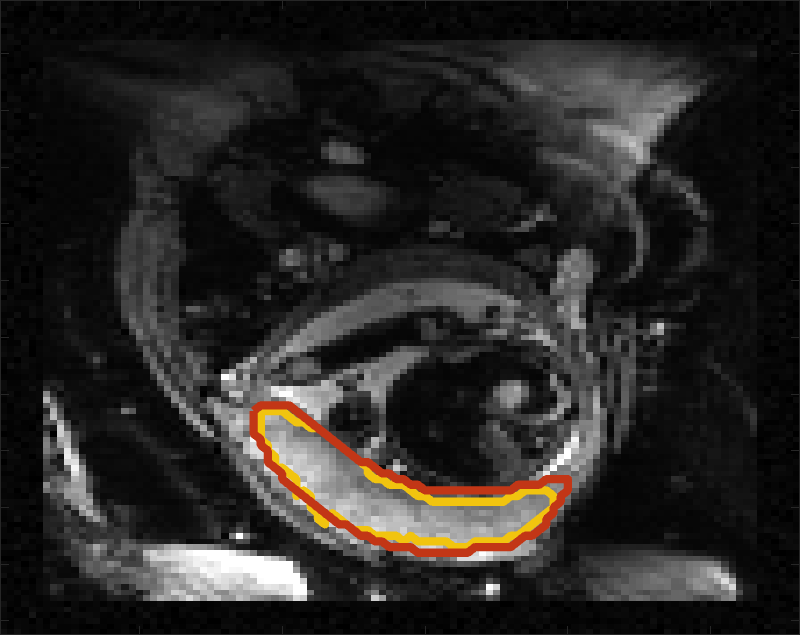} & \includegraphics[width=.19\linewidth]{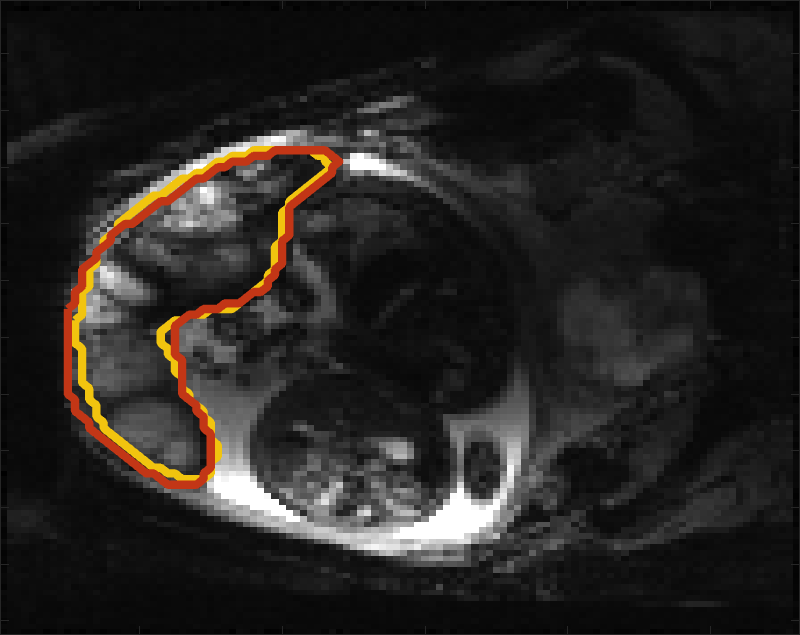} \\
    $0.67$ & $0.76$ & $0.81$ & $0.84$ & $0.91$
    \end{tabular}
    \caption{Example predictions on $5$ subjects from the test set. Ground truth segmentations are shown in yellow and predictions in red. Dice scores are indicated below each column. Two slices are shown for each subject, spaced $18$mm apart.}
    \label{fig:test-results}
\end{figure}

Our model performs consistently well in the normoxic and hyperoxic phases. For the $5$ subjects with ground truth segmentations in both the normoxic and hyperoxic phase, we achieve a mean absolute difference between predictions in normoxia and hyperoxia of $0.026\pm0.02$ Dice, $5.69\pm2.33$mm HD95, $0.75\pm0.46$mm ASSD, and $0.06\pm0.04$ relative BOLD error. These results suggest that our model is robust to contrast changes in the placenta resulting from maternal hyperoxia, and can be used in studies quantifying oxygen transport in the organ. A larger number of subjects are needed to assess statistical significance.

Fig~\ref{fig:test-results} compares the predicted label maps with ground truth on $5$ subjects with increasing Dice scores using the BW-CE model. The model accurately identifies the location of the placenta, but in the worst cases misses boundary details. 



\subsubsection{BOLD Time Series Evaluation}

Table~\ref{tab:consistency-results} presents statistics of the consistency between predicted label maps in consecutive volumes of the MRI time series. Predictions are highly consistent, achieving a Dice of $0.92\pm0.02$.  The small differences between the relative mean-BOLD values suggest these produced segmentations may be suitable for research studies assessing placental function.

\begin{figure}[t]
    \centering
    \includegraphics[width=0.84\textwidth]{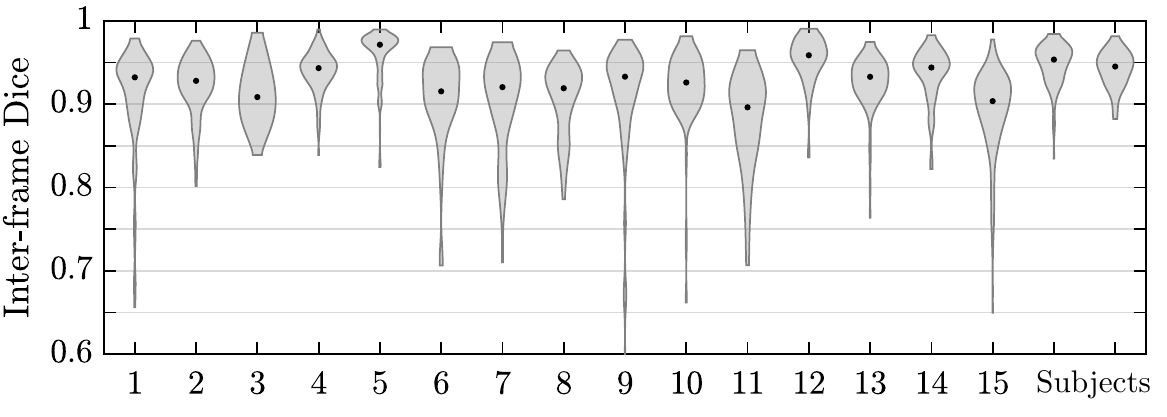}
    \caption{Per-subject density distributions of Dice scores between consecutive predictions in BOLD MRI time series.  Dots inside  distributions indicate the median.}
    \label{fig:consecutive-bold}
\end{figure}

Fig.~\ref{fig:consecutive-bold} presents distributions of Dice score between predicted label maps of consecutive frames in the BOLD time series. Distributions have high medians ($>0.9$) for all but one case, with wide density at high Dice scores ($>0.9$. Dice differences are highly affected by fetal and maternal motion that causes placental deformation. We visually verified that modest drops in Dice ($<0.9$) were mainly due to fetal motion, but large drops (Dice $<0.7$) resulted from errors in the produced label maps.

\begin{table}[b]
\caption{Consistency of predictions in the BOLD time series produced by our best-performing 3D U-Net model (trained using the BW-CE loss function).}
\label{tab:consistency-results}
    \includegraphics[width=\linewidth]{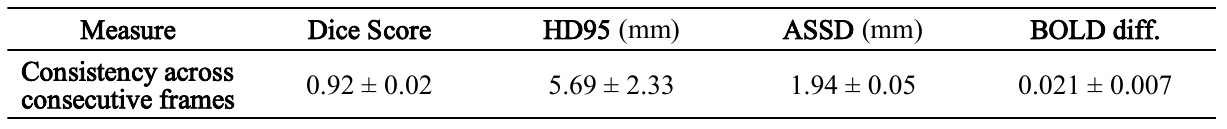}
\end{table}

Automatic segmentation of each volume in BOLD MRI time series is advantageous as it can enable whole-organ spatiotemporal analysis without requiring inter-volume motion correction or registration, which may fail under the presence of large motion. 
We illustrate one possible application by investigating the percentage increase in BOLD signal in response to maternal hyperoxia. We calculate the percentage increase over the baseline period: $\Delta b =  \vert b_{H}-b_{N}\vert/b_{N}$, where $b_{N}$ denotes the mean BOLD signal over the baseline period, and $b_{H}$ denotes the mean of the signal in the last $10$ frames of the hyperoxic period.

\begin{figure}[t]
\scriptsize
    \centering
    \begin{tabular}{p{0.5\textwidth}p{0.5\textwidth}}
    \includegraphics[width=0.92\linewidth]{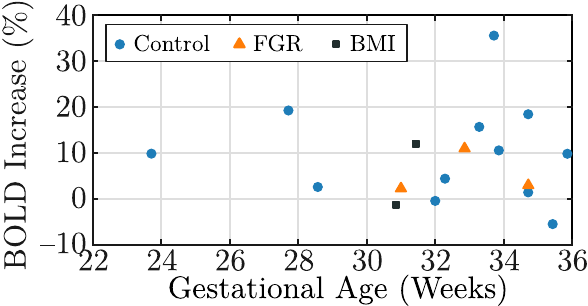} &
    \includegraphics[width=0.92\linewidth]{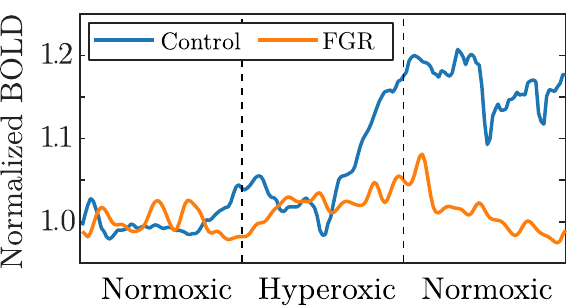} 
    \end{tabular}
    \caption{Example application using our model's produced placenta segmentations in BOLD time series to characterize oxygenation response from maternal hyperoxia. Left: observed increase relative to baseline for the test set. Right: example time series for one singleton control (GA=33wk2d, Dice=$0.84$, $\Delta b=15.7\%$) and one singleton FGR subject (GA=34wk5d, Dice=$0.84$, $\Delta b=2.9\%$).}
    \label{fig:example-application}
\end{figure}

Fig.~\ref{fig:example-application} shows a scatter plot of the hyperoxia response for all subjects in the test set and two examples of the BOLD signal time course in the produced placenta segmentation label maps. In the control subjects (N=$12$), we observe an increase of $10.2\pm 11.1\%$. The observed increase for the healthy controls is consistent with previous studies that demonstrated an increase of $12.6\pm5.4\%$ (N=$21$)~\cite{sorensen2015placental} and from $5\%$ to $20\%$ throughout gestation (N=$49$)~\cite{sinding2018placental}. 


\section{Discussion and Conclusion}
We developed a model to automatically segment placental scans in BOLD MRI and achieve close matching to ground truth labels with consistent performance in predicting volumes in both the normoxic and hyperoxic phases. Key to our model development is a boundary-weighted loss function and training with labeled volumes obtained at different oxygenation phases in the BOLD MRI time series. 

Segmenting each volume in the BOLD MRI time series can be advantageous for clinical research assessing whole-organ changes as it eliminates the need for registration. Registration algorithms are affected by fetal motion and may require discarding a significant number of volumes~\cite{turk2017spatiotemporal,you2015robust}, potentially losing important signal information. 
We illustrate one possible study in assessing placental response during hyperoxia, observing an increase in signal intensity consistent with prior work. However, our cohort is limited, and several factors, including maternal position, gestational age, and contractions are covariates not considered.

Registration however is advantageous for localized analysis~\cite{turk2017spatiotemporal}, and solely relying on segmentation would only permit quantifying whole-organ signal changes, for example mean T2* or mean BOLD increase. Placental segmentations can be incorporated into registration methods as spatial priors to improve registration results. Future work will investigate joint segmentation-registration models.

We assessed the consistency of predictions in BOLD MRI time series using our model, and achieved highly consistent predictions (Dice $=0.92$). For many subjects, we observed modest drops in Dice ($<0.9$), which were often due to fetal motion displacing the placenta. However, in a small number of cases, we observed large drops (Dice $<0.7$) that we visually verified were caused by segmentation error. Since we apply the model to each volume in the time series independently, imaging artifacts, such as intensity and geometric artifacts, can affect the predicted segmentations. In future work, we will investigate incorporating temporal consistency between consecutive volumes. We will also investigate applying test-time augmentation on image intensity as this has been shown to reduce uncertainty and improve segmentation robustness~\cite{WANG2019testtime}.

Key to our model performance was maximizing data variability by having manually segmented volumes at different points in the BOLD MRI series. Future work will investigate semi-supervised learning to incorporate all unlabeled volumes. As there are often in the order of 100 unlabeled volumes in each BOLD time series, these approaches can more accurately capture the rapid signal changes resulting from fetal motion and maternal oxygenation. 

Future directions of this work will investigate  oxygenation dynamics in the placenta. Segmentation of the time series can be used to derive T2* maps and perform whole-organ signal comparisons between differing population groups, thereby enabling quantitative analysis of placental function with the ultimate goal of developing biomarkers of placental and fetal health.

\section{Acknowledgments}
This work was supported in part by NIH NIBIB NAC P41EB015902, NIH NICHD R01HD100009, R01EB032708, R21HD106553, MIT-IBM Watson AI Lab, NSERC PGS D, NSF GRFP, and MathWorks Fellowship.




%
%
%
\bibliographystyle{assets/splncs04_official}
\bibliography{main}
\end{document}